%% file: AyacheBertrandArxiv.tex
\documentclass{svmult}

\usepackage{graphicx}

\input epsf
\usepackage{multicol}
\usepackage[bottom]{footmisc}

\def\a{\alpha}

\def\th{\theta}

\renewcommand{\Box}{\hfill\rule{0.25cm}{0.25cm}} 

\def\I{{\rm 1 \hskip-2.9truept l}}

\def\rit{\hbox{\it I\hskip -2pt R}}
\def\nit{\hbox{\it I\hskip -2pt N}}

\def\zit{\hbox{\it Z\hskip -4pt Z}}

\def\eps{\varepsilon}

\def\E{{\mathbb E}}

\begin{document}

\title*{A process very similar to multifractional Brownian motion}
\author{Antoine Ayache\inst{1}\and
Pierre R. Bertrand\inst{2}}
\institute{UMR CNRS 8524, Laboratoire Paul Painlev\'e, B\^at. M2, Universit\'e Lille 1, 59655 Villeneuve d'Ascq, FRANCE
\texttt{Antoine.Ayache@math.univ-lille1.fr}
\and INRIA Saclay and Universit\'e Clermont-Ferrand, UMR CNRS 6620
 \texttt{Pierre.Bertrand@inria.fr}}
%
%
\maketitle
{\small Abstract : 
Multifractional Brownian motion (mBm), denoted here by $X$, is one of the paradigmatic examples of a continuous Gaussian process whose pointwise H\"older exponent depends on the location. Recall that $X$ can be obtained (see e.g. \cite{bjr97,at05}) by replacing the constant Hurst parameter $H$ in the standard wavelet series representation of
fractional Brownian motion (fBm) by a smooth function $H(\cdot)$ depending on
the time variable $t$. Another natural idea (see \cite{bbci00}) which allows
to construct a continuous Gaussian process, denoted by $Z$, whose pointwise
H\"older exponent does not remain constant all along its trajectory, consists
in substituting $H(k/2^j)$ to $H$ in each term of index $(j,k)$ of the
standard wavelet series representation of fBm. The main goal of our article is to show that $X$ and $Z$  only differ by a process $R$ which is smoother than them; this means that they are very similar from a fractal geometry point of view.}
\par
\bigskip
\noindent
{\bf Keywords:} Fractional Brownian motion, wavelet series expansions,
multifractional Brownian motion, H\H{o}lder regularity.

\section{Introduction and statement of the main results}
\label{sec:1}
Throughout this article we denote by $H(\cdot)$ an arbitrary function defined
on the real line and with values in an arbitrary fixed compact interval
$[a,b]\subset(0,1)$. We will always assume that on each compact ${\cal
  K}\subset\rit$, $H(\cdot)$ satisfies a uniform H\H{o}lder condition of order
$\beta>b$ i.e. there is a constant $c_1>0$ (which a priori depends on ${\cal
  K}$) such that for every $t_1,t_2\in {\cal K}$ one has,
\begin{equation}
\label{eq:HoldH}
|H(t_1)-H(t_2)|\le c_1 |t_1-t_2|^\beta;
\end{equation}
typically $H(\cdot)$ is a Lipschitz function over $\rit$. We
will also assume that $a=\inf\{H(t)\,:\,t\in\rit\}$ and
$b=\sup\{H(t)\,:\,t\in\rit\}$. Recall that multifractional Brownian motion (mBm) of functional parameter $H(\cdot)$, which we denote by $X=\{X(t):\,\,t\in\rit\}$, is the continuous and nowhere differentiable Gaussian process obtained by replacing the Hurst parameter in the harmonizable representation of fractional Brownian motion (fBm) by the function $H(\cdot)$. That is, the process $X$ can be represented for each $t\in\rit$ as the following stochastic integral
\begin{equation}
\label{eq:defmbm}
X(t)=\int_{\rit}\frac{e^{it\xi}-1}{|\xi|^{H(t)+1/2}}\,d\widehat{W}(\xi),
\end{equation}
where $d\widehat{W}$ is ``the Fourier transform'' of the real-valued white-noise $dW$ in the sense that for any function $f\in L^2(\rit)$ one has a.s.
\begin{equation}
\label{eq:whitenoise}
\int_{\rit}f(x)\,dW(x)=\int_{\rit}\widehat{f}(\xi)\,d\widehat{W}(\xi).
\end{equation}
Observe that (\ref{eq:whitenoise}) implies that (see \cite{c99,st06}) the following equality holds a.s. for every $t$, to within a deterministic smooth bounded and non-vanishing deterministic function,
$$
\int_{\rit}\frac{e^{it\xi}-1}{|\xi|^{H(t)+1/2}}\,d\widehat{W}(\xi)=\int_{\rit}\Big\{|t-s|^{H(t)-1/2}-|s|^{H(t)-1/2}\Big\}\,dW(s).
$$
Therefore $X$ is a real-valued process. MBm was introduced independently in
\cite{plv95} and \cite{bjr97} and since then there is an increasing interest
in the study of multifractional processes, we refer for instance to
\cite{fal,s08} for two excellent quite recent articles on this topic. The main three features of mBm are the following:
\begin{itemize}
\item[(a)] $X$ reduces to a fBm when the function $H(\cdot)$ is constant.
\item[(b)] Unlike to fBm, $\a_X=\{\a_X (t):\,\,t\in\rit\}$ the pointwise H\"older exponent of $X$ may depend on the location and can be prescribed via the functional parameter $H(\cdot)$; in fact one has (see \cite{plv95,bjr97,at05,ajt07}) a.s. for each $t$,
\begin{equation}
\label{eq:phembm}
\a_X (t)=H(t).
\end{equation}
Recall that $\a_X $ the pointwise H\"older exponent of an arbitrary continuous and nowhere differentiable process $X$, is defined, for each $t\in\rit$, as
\begin{equation}
\label{eq:phe}
\a_X (t)=\sup\left\{\a\in\rit_+:\,\,\limsup_{h\rightarrow 0}\frac{|X(t+h)-X(t)|}{|h|^\a}=0\right\}.
\end{equation}
\item[(c)] At any point $t\in\rit$, there is an fBm of Hurst parameter $H(t)$, which is tangent to mBm \cite{bjr97,f02,f03} i.e. for each sequence $(\rho_n)$ of positive real numbers converging to $0$, one has,
\begin{equation}
\label{eq:tangentX}
\lim_{n\rightarrow\infty}\mbox{law}\left\{\frac{X(t+\rho_n u)-X(t)}{\rho_n^{H(t)}}:\,\,u\in\rit\right\}=\mbox{law}\{B_{H(t)}(u):\,\,u\in\rit\},
\end{equation}
where the convergence holds in distribution for the topology of uniform convergence on compact sets.
\end{itemize}
The main goal of our article is to give a natural wavelet construction of a continuous and nowhere differentiable Gaussian process $\displaystyle Z=\{Z(t)\}_{t\in\rit}$ which has the same features $(a)$, $(b)$ and $(c)$ as mBm $X$ and which differs from it by a smoother stochastic process $R=\{R(t):\,\,t\in\rit\}$ (see Theorem \ref{thm:main1}).

In order to be able to construct $Z$, first we need to introduce some notation. In what follows we denote by $\{2^{j/2}\psi(2^j x-k):\,\,(j,k)\in\zit^{\,2}\}$ a Lemari\'e-Meyer wavelet basis of $L^2(\rit)$ \cite{lm86} and we define   $\Psi$ to be the function , for each $(x,\th)\in\rit\times\rit$,
\begin{equation}
\label{eq:defpsi}
\Psi(x,\th)=\int_{\rit}e^{ix\xi}\frac{\widehat{\psi}(\xi)}{|\xi|^{\th+1/2}}\,d\xi.
\end{equation}
 By using the fact that $\widehat{\psi}$ is a compactly supported $C^\infty$
 function vanishing on a neighborhood of the origin, it follows that $\Psi$ is
 a well-defined $C^\infty$ function satisfying for any $(l,m,n)\in\nit^{\,3}$ with $l\ge 2$, the following localization property
(see \cite{at05} for a proof),
\begin{equation}
\label{eq:uniflocpsi}
c_2= \sup_{\th\in[a,b],\, x\in \rit}(2+|x|)^{\ell}|(\partial_{x}^m\partial_{\th}^n\Psi)(x,\th)|<\infty,
\end{equation}
where $\partial_{x}^m\partial_{\th}^n\Psi$ denotes the function obtained by differentiating the function $\Psi$, $n$ times with respect to the variable $\th$ and $m$ times with respect to the variable $x$. For convenience, let us introduce the Gaussian field $B=\{B(t,\th):\,\,(t,\th)\in \rit\times (0,1)\}$ defined for each $(t,\th)\in \rit\times (0,1)$ as
\begin{equation}
\label{eq:defB}
B(t,\th)=\int_{\rit}\frac{e^{it\xi}-1}{|\xi|^{\th+1/2}}\,d\widehat{W}(\xi).
\end{equation}
Observe that for every fixed $\th$, the Gaussian process $B(\cdot,\th)$ is an fBm of Hurst parameter $\th$ on the real line. Also observe that mBm $X$ satisfies for each $t\in\rit$,
\begin{equation}
\label{eq:connXB}
X(t)=B(t,H(t)).
\end{equation}
By expanding for every fixed $(t,\th)$, the kernel function $\displaystyle \xi\mapsto\frac{e^{it\xi}-1}{|\xi|^{\th+1/2}}$ in the orthonormal basis of $L^2(\rit)$, $\displaystyle \{2^{-j/2}(2\pi)^{1/2}e^{i2^{-j}k\xi}\widehat{\psi}(-2^{-j}\xi):\,\, (j,k)\in \zit^{\,2}\}$ and by using the isometry property of the stochastic integral in (\ref{eq:defB}), it follows that
\begin{equation}
\label{eq:wavB}
B(t,\th)=\sum_{j=-\infty}^{\infty}\sum_{k=-\infty}^{\infty} 2^{-j\th}\eps_{j,k}\,
\Big\{\Psi(2^j t-k,\th)-\Psi(-k,\th)\Big\},
\end{equation}
where $\{\eps_{j,k}:\,\,(j,k)\in\zit^{\,2}\}$ is a sequence of independent ${\cal N}(0,1)$ Gaussian random variables and where the series is, for every fixed $(t,\th)$, convergent in $L^2(\Omega)$; throughout this article $\Omega$ denotes the underlying probability space.
In fact this series is also convergent in a much stronger sense, see part $(i)$ of
the following remark.
\begin{remark}
\label{rem:fieldB}
The field $B$ has already been introduced and studied in \cite{at05}; we recall some of its useful properties:
\begin{itemize}
\item[(i)] The series in (\ref{eq:wavB}) is a.s. uniformly convergent in
  $(t,\th)$ on each compact subset of $\rit\times (0,1)$, so $B$ is a
  continuous Gaussian field. Moreover, combining (\ref{eq:connXB}) and (\ref{eq:wavB}), we deduce the following wavelet expansion of mBm,
\begin{equation}
\label{eq:wavmBm}
X(t)=\sum_{j=-\infty}^{\infty}\sum_{k=-\infty}^{\infty} 2^{-jH(t)}\eps_{j,k}
\Big\{\Psi(2^j t-k,H(t))-\Psi(-k,H(t))\Big\}.
\end{equation}

\item[(ii)] The low frequency component of $B$, namely the field $\dot{B}=\{\dot{B}(t,\th):\,\,(t,\th)\in\rit\times (0,1)\}$ defined for all $(t,\th)\in\rit\times (0,1)$ as
\begin{equation}
\label{eq:lfB}
\dot{B}(t,\th)=\sum_{j=-\infty}^{-1}\sum_{k=-\infty}^{\infty} 2^{-j\th}\eps_{j,k}\,
\Big\{\Psi(2^j t-k,\th)-\Psi(-k,\th)\Big\},
\end{equation}
is a $C^\infty$ Gaussian field. Therefore (\ref{eq:HoldH}) and (\ref{eq:connXB}) imply that the low frequency component of the mBm $X$, namely the Gaussian process $\dot{X}=\{\dot{X}(t)\}_{t\in\rit}$ defined for each $t\in\rit$ as
\begin{equation}
\label{eq:lfX}
\dot{X}(t)=\sum_{j=-\infty}^{-1}\sum_{k=-\infty}^{\infty} 2^{-jH(t)}\eps_{j,k}\,
\Big\{\Psi(2^j t-k,H(t))-\Psi(-k,H(t))\Big\},
\end{equation}
satisfies a uniform H\"older condition of order $\beta$ on each compact subset ${\cal K}$ of $\rit$. Thus, in view of $(b)$ and the assumption $b<\beta$, the pointwise H\"older exponent of $X$ is only determined by its high frequency component, namely the continuous Gaussian process $\ddot{X}=\{\ddot{X}(t)\}_{t\in\rit}$ defined for each $t\in\rit$ as
\begin{equation}
\label{eq:hfX}
\ddot{X}(t)=\sum_{j=0}^{+\infty}\sum_{k=-\infty}^{\infty} 2^{-jH(t)}\eps_{j,k}\,
\Big\{\Psi(2^j t-k,H(t))-\Psi(-k,H(t))\Big\}.
\end{equation}
\end{itemize}
\end{remark}

\begin{definition}
\label{def:procZ}
The process $Z=\{Z(t):\,\,t\in\rit\}$ is defined for each $t\in\rit$ as
\begin{equation}
\label{eq:defZ}
Z(t)=\sum_{j=-\infty}^{\infty}\sum_{k=-\infty}^{\infty} 2^{-jH(k/2^j)}\eps_{j,k}\,
\Big\{\Psi(2^j t-k,H(k/2^j))-\Psi(-k,H(k/2^j))\Big\}.
\end{equation}
\end{definition}
In view of (\ref{eq:wavB}) it is clear that the process $Z$ reduces to a fBm when the function $H(\cdot)$ is constant; this means that the process $Z$ has the same feature $(a)$ as mBm.
\begin{remark}
\label{rem:welldZ}
Using the same technics as in \cite{at05} one can show that:
\begin{itemize}
\item[(i)] The series in (\ref{eq:defZ}) is a.s. uniformly convergent in $t$ on each compact interval of $\rit$; therefore $Z$ is a well-defined continuous Gaussian process.
\item[(ii)] The low frequency component of the process $Z$, namely the process $\dot{Z}=\{\dot{Z}(t):\,\,t\in\rit\}$ defined for all $t\in\rit$ as
\begin{equation}
\label{eq:defZdot}
\dot{Z}(t)=\sum_{j=-\infty}^{-1}\sum_{k=-\infty}^{\infty} 2^{-jH(k/2^j)}\eps_{j,k}\,
\Big\{\Psi(2^j t-k,H(k/2^j))-\Psi(-k,H(k/2^j))\Big\},
\end{equation}
is a $C^\infty$ Gaussian process. The pointwise H\"older exponent of $Z$ is therefore only determined by its high frequency component, namely the continuous Gaussian process $\ddot{Z}=\{\ddot{Z}(t):\,\,t\in\rit\}$ defined for all $t\in\rit$ as
\begin{equation}
\label{eq:defZddot}
\ddot{Z}(t)=\sum_{j=0}^{+\infty}\sum_{k=-\infty}^{\infty} 2^{-jH(k/2^j)}\eps_{j,k}\,
\Big\{\Psi(2^j t-k,H(k/2^j))-\Psi(-k,H(k/2^j))\Big\}.
\end{equation}
\end{itemize}
It is worth noticing that if one replaces in (\ref{eq:defZddot}) the H\"older function $H(\cdot)$ by a step function then one recovers the step fractional Brownian motion which has been studied in \cite{bbci00,ablv07}.
\end{remark}

\begin{figure}[htbp]
\label{fig:f1}
\begin{center}
\includegraphics[width=6cm,height=6cm]{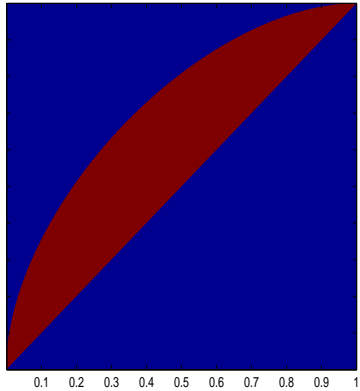}
\caption{the region $\mathcal{D}$ in the unit cube satisfying (\ref{eq:condab})}
\end{center}
\end{figure}

Let us now state our main result.
\begin{theorem}
\label{thm:main1}
Let $R=\{R(t):\,\,t\in\rit\}$ be the process defined for any $t\in\rit$ as
\begin{equation}
\label{eq:defR}
R(t)=Z(t)-X(t).
\end{equation}
Let $\cal K$ be a compact interval included in $\rit$.
Then, if $a$ and $b$  satisfy the following condition:
\begin{equation}
\label{eq:condab}
1-b>(1-a)(1-ab^{-1}),
\end{equation}
there exists an exponent $d\in (b,1]$, such that the process $R$ satisfies a uniform H\"older condition of order $d$ on  $\cal K$. More precisely, there is $\Omega^*$ an event of probability $1$, such that, for all $\omega\in\Omega^*$ and for each $(t_0,t_1)\in {\cal K}^2$, one has
\begin{equation}
\label{eq1:main1}
|R(t_1,\omega)-R(t_0,\omega)|\le C_1 (\omega)|t_1-t_0|^d,
\end{equation}
where $C_1$ is a nonnegative random variable of finite moment of every order only depending on $\Omega^*$ and $\cal K$.
\end{theorem}
\begin{remark}
\label{rem:cond19}
We do not know whether Theorem \ref{thm:main1} remains valid when Condition
(\ref{eq:condab}) does not hold. Figure \ref{fig:f1} below indicates the
region $\mathcal{D}$ in the unit cube satisfying (\ref{eq:condab}).
\end{remark}

Thanks to the previous theorem we can obtain the following result which shows that $Z$ and $X$ are very similar from a fractal geometry point of view.
\begin{corollary}
\label{cor:main2}
Assume that $a$ and $b$ satisfy (\ref{eq:condab}), then the process $Z$ has the same features $(a)$, $(b)$ and $(c)$ as mBm.
\end{corollary}

Throughout this article, we use $[x]$ to denote the integer part of a real number $x$. Positive deterministic constants will be numbered as $c_1,c_2,\ldots$ while positive random constants will be numbered as $C_1, C_2,\ldots$.

\section{The main ideas of the proofs}
\label{sec:proofs}
In the reminder of our article we always assume that Condition (\ref{eq:condab}) is satistfied and that $\mbox{diam}({\cal K}):=\sup\{|u-v|:\,(u,v)\in {\cal K}\}\le 1/4$. Also notice that we will frequently make use of the inequality
\begin{equation}
\label{bound:log3xy}
\log(3+x+y) \le \log(3+x) \times\log(3+y)\qquad \mathrm{for\; all}\; (x,y) \in \rit_+^2.
\end{equation}
Let us now present the main ideas behind the proof of Theorem \ref{thm:main1}.
Firstly we need to state the following lemma which allows to conveniently
bound the random variables $\eps_{j,k}$. It is a classical result we refer for example to \cite{mst99} or \cite{at03} for its proof.
\begin{lemma}
\label{lem:boundeps}
\cite{mst99,at03} There are an event $\Omega^*$ of probability $1$ and a nonnegative random variable $C_2$ of finite moment of every order such the inequality
\begin{equation}
\label{eq:boundeps}
|\eps_{j,k}(\omega)|\le C_2(\omega)\sqrt{\log(3+|j|+|k|)},
\end{equation}
holds for all $\omega\in\Omega^*$ and $j,k\in\zit$.
\end{lemma}
{\it Proof of Theorem \ref{thm:main1}.} In view of Remark \ref{rem:fieldB} $(ii)$ and of Remark \ref{rem:welldZ} it is sufficient to prove that Theorem \ref{thm:main1} holds when the process $R$ is replaced by its high frequency component, namely the process $\ddot{R}=\{\ddot{R}(t):\,\,t\in\rit\}$ defined for each $t\in\rit$ as
\begin{equation}
\label{eq:defRddot}
\ddot{R}(t)=\ddot{Z}(t)-\ddot{X}(t).
\end{equation}
Let $g_{j,k}$ be the function defined on $\rit\times \rit$ by
\begin{equation}
\label{eq:defgjk}
g_{j,k}(t,\th)=2^{-j\th}\Big\{\Psi(2^j t-k,\th)-\Psi(-k,\th)\Big\}.
\end{equation}
It follows from (\ref{eq:defRddot}), (\ref{eq:hfX}), (\ref{eq:defZddot}), (\ref{eq:defgjk}) and (\ref{eq:boundeps}) that for any $\omega\in\Omega^*$,
\begin{eqnarray}
\label{eq2:main1}
&&|\ddot{R}(t_1,\omega)-\ddot{R}(t_0,\omega)| \le C_2 (\omega)\sum_{j=0}^{+\infty}\sum_{k=-\infty}^{+\infty}\sqrt{\log(3+j+|k|)}
\\\nonumber
&&\hspace{0.3cm}
\times \Big |g_{j,k}(t_1,H(k/2^j))-g_{j,k}(t_0,H(k/2^j))-g_{j,k}(t_1,H(t_1))+g_{j,k}(t_0,H(t_0))\Big|.
\end{eqnarray}
Next, we expand the term $g_{j,k}\big(t_i, H(\tau)\big)$ with $i=0$ or $1$ and $\tau=t_1$ or $k/2^j$  with respect to the second variable in the neighborhood of $H(t_0)$.
Indeed, since the function $\Psi$ is $\mathcal{C}^\infty$, the functions $g_{j,k}$ are also $\mathcal{C}^\infty$; thus
 we can use  Taylor-Lagrange Formula of order $1$ with an integral reminder and we get
\begin{eqnarray}
\label{eq:tf1}
g_{j,k}(t_1,H(t_1))&=& g_{j,k}(t_1,H(t_0))+ (H(t_1)-H(t_0))(\partial_\th g_{j,k})(t_1,H(t_0))\\
\nonumber
&&\hspace{-2cm}+(H(t_1)-H(t_0))^2\int_0^1 (1-\tau)(\partial_\th ^2 g_{j,k})(t_1,H(t_0)+\tau (H(t_1)-H(t_0)))\,d\tau,
\end{eqnarray}
\begin{eqnarray}
\label{eq:tf2}
g_{j,k}(t_0,H(k/2^j))&=& g_{j,k}(t_0,H(t_0))+ (H(k/2^j)-H(t_0))(\partial_\th g_{j,k})(t_0,H(t_0))\\
\nonumber
&&\hspace{-2.8cm}+(H(k/2^j)-H(t_0))^2\int_0^1 (1-\tau)(\partial_\th ^2 g_{j,k})(t_0,H(t_0)+\tau (H(k/2^j)-H(t_0)))\,d\tau,
\end{eqnarray}
and
\begin{eqnarray}
\label{eq:tf3}
g_{j,k}(t_1,H(k/2^j))&=& g_{j,k}(t_1,H(t_0))+ (H(k/2^j)-H(t_0))(\partial_\th g_{j,k})(t_1,H(t_0))\\
\nonumber
&&\hspace{-2.8cm}+(H(k/2^j)-H(t_0))^2\int_0^1 (1-\tau)(\partial_\th ^2 g_{j,k})(t_1,H(t_0)+\tau (H(k/2^j)-H(t_0)))\,d\tau.
\end{eqnarray}
By adding or subtracting relations (\ref{eq:tf1}), (\ref{eq:tf2}) and (\ref{eq:tf3}) the constant terms disappear and we get the following upper bound
\begin{eqnarray}\label{bound:gjk}
&& \Big |g_{j,k}(t_1,H(k/2^j))-g_{j,k}(t_0,H(k/2^j))-g_{j,k}(t_1,H(t_1))+g_{j,k}(t_0,H(t_0))\Big|\\
\nonumber
&& \le |H(t_1)-H(t_0)|\Big|(\partial_\th g_{j,k})(t_1,H(t_0))\Big|\\
\nonumber
&& +\big|H(t_1)-H(t_0)\big|^2\int_0^1 (1-\tau)\Big |(\partial_\th ^2 g_{j,k})(t_1,H(t_0)+\tau (H(t_1)-H(t_0)))\Big|\,d\tau\\
\nonumber
&&+|H(k/2^j)-H(t_0)|\Big |(\partial_\th g_{j,k})(t_1,H(t_0))-(\partial_\th g_{j,k})(t_0,H(t_0))\Big |\\
\nonumber
&&+\big|H(k/2^j)-H(t_0)\big|^2 \int_0^1 (1-\tau)\Big | (\partial_\th ^2 g_{j,k})(t_1,H(t_0)+\tau (H(k/2^j)-H(t_0)))\\
\nonumber
&& \hspace{4.5cm}-(\partial_\th ^2 g_{j,k})(t_0,H(t_0)+\tau (H(k/2^j)-H(t_0)))\Big|\,d\tau.
\end{eqnarray}
Then, we substitute the previous bound (\ref{bound:gjk}) into the inequality (\ref{eq2:main1}). We stress that the quantities
$|H(t_1)-H(t_0)|$ and $\big|H(t_1)-H(t_0)\big|^2$ can be factorized outside the sum whereas the quantities
$|H(k/2^j)-H(t_0)|$ and $\big|H(k/2^j)-H(t_0)\big|^2 $ remain inside the sum. We obtain
\begin{eqnarray}
\label{eq2:main2}
\nonumber
|\ddot{R}(t_1,\omega)-\ddot{R}(t_0,\omega)| &\le& C_2 (\omega)|H(t_1)-H(t_0)|
\\
\nonumber
\hspace{-1cm}&&\times
\Big\{\sum_{j=0}^{+\infty}\sum_{k=-\infty}^{+\infty}\sqrt{\log(3+j+|k|)}\times\Big|(\partial_\th g_{j,k})(t_1,H(t_0))\Big|\Big\}
\\ \nonumber
&&\hspace{-2cm}+\;C_2 (\omega)|H(t_1)-H(t_0)|^2\times\Big\{
\sum_{j=0}^{+\infty}\sum_{k=-\infty}^{+\infty}\sqrt{\log(3+j+|k|)}
\\ \nonumber
&&
\hspace{-0.5cm}
\times\int_0^1 (1-\tau)\Big |(\partial_\th ^2 g_{j,k})(t_1,H(t_0)+\tau (H(t_1)-H(t_0)))\Big|\,d\tau\Big\}
\\ \nonumber&&\hspace{-2cm}+\; C_2 (\omega)
 \times
\Big\{\sum_{j=0}^{+\infty}\sum_{k=-\infty}^{+\infty}\sqrt{\log(3+j+|k|)}|H(k/2^j)-H(t_0)|
\\ \nonumber
&&
\times\Big |(\partial_\th g_{j,k})(t_1,H(t_0))-(\partial_\th g_{j,k})(t_0,H(t_0))\Big |\Big\}
\\\nonumber
&&\hspace{-2cm}+\;C_2 (\omega)
 \times
\Big\{\sum_{j=0}^{+\infty}\sum_{k=-\infty}^{+\infty}\sqrt{\log(3+j+|k|)}|H(k/2^j)-H(t_0)|^2
\\ \nonumber
&&
 \times\hspace{-0.1cm}\int_0^1 (1-\tau)\Big | (\partial_\th ^2 g_{j,k})(t_1,H(t_0)+\tau (H(k/2^j)\hspace{-0.1cm}-\hspace{-0.1cm}H(t_0)))
 \\ \nonumber
 &&\hspace{0.8cm}
 -(\partial_\th ^2 g_{j,k})(t_0,H(t_0)+\tau (H(k/2^j)-H(t_0)))\Big|\,d\tau\Big\}.
\end{eqnarray}
Then using the following two lemmas whose proofs will be given soon, we get that
\begin{eqnarray}
\label{eq3:main1}
 |\ddot{R}(t_1,\omega)-\hspace{-0.1cm}\ddot{R}(t_0,\omega)|
& \le& C_2(\omega)\Big\{|H(t_1)-\hspace{-0.1cm}H(t_0)|{\cal A}_1({\cal K};a,b)+\dots\\
\nonumber
&&+\;\big|H(t_1)- H(t_0)\big|^2 {\cal A}_2({\cal K};a,b)+\dots\\
\nonumber
&&+\; |t_1-t_0|^{d_1}  {\cal G}_1({\mathcal K};a,b,d_1)+|t_1-t_0|^{d_2} {\cal G}_2({\mathcal K};a,b,d_2)\Big\}.
\end{eqnarray}
Finally, in view of (\ref{eq:HoldH}) the latter inequality implies that Theorem \ref{thm:main1} holds.
$\Box $
\begin{lemma}
\label{lem:boundg}
For every integer $n\ge 0$ and $(t,\th)\in\rit\times (0,+\infty)$ one sets
\begin{equation}
\label{eq:defA}
A_n (t,\th):=\sum_{j=0}^{+\infty}\sum_{k=-\infty}^{+\infty}|(\partial_{\th}^n g_{j,k}(t,\th)|\sqrt{\log(3+j+|k|)}.
\end{equation}
Then one has
\begin{equation}
\label{eq1:boundg}
{\cal A}_n({\cal K};a,b):=\sup\Big\{A_n (t,\th):\,\,(t,\th)\in {\cal K} \times [a,b]\Big\}<\infty.
\end{equation}
\end{lemma}

\begin{lemma}
\label{lem:d1boundg}
For every integer $n\ge 1$ and $(t_0, t_1, \th)\in\rit^2\times  (0,+\infty)$ one sets
\begin{eqnarray*}
\nonumber
G_n(t_0, t_1, \th)&:=&\sum_{j=0}^{+\infty}\sum_{k=-\infty}^{+\infty} |H(k/2^j)-H(t_0)|^n \times \sqrt{\log(3+j+|k|)}
\\
&&\hspace{1.5cm}\times \Big |(\partial_{\th}^n g_{j,k})(2^j t_1-k,\th)-(\partial_{\th}^n g_{j,k})(2^jt_0-k,\th)\Big|.
\end{eqnarray*}
Then, for every integer $n\ge 1$, there is an exponent $d_n\in(b,1]$ such that
\begin{eqnarray}
\label{eq1:d1boundg}
{\cal G}_n({\mathcal K};a,d_n):=\sup_{(t_0,t_1,\th)\in {\cal K}^2\times [a,b]} |t_1-t_0|^{-d_n} G_n(t_0, t_1,\th)<\infty.
\end{eqnarray}
\end{lemma}
\par
\medskip
{\it Proof of Lemma \ref{lem:boundg}.} From Lemma \ref{lem:leibniz} given in
next section, one can deduce
\begin{eqnarray}\label{maj:An}
A_n(t,\th)&\le& \sum_{p=0}^n C_{n}^p |\log 2|^p\sum_{j=0}^{+\infty}\sum_{k=-\infty}^{+\infty}
j^p 2^{-j\th}\sqrt{\log (3+j+|k|)}
\\
\nonumber
&&\hspace{3cm}\times\Big\{\big|(\partial_{\th}^{n-p}\Psi)(2^j t-k,\th)\big|+ \big|(\partial_{\th}^{n-p}\Psi)(-k,\th)\big|\Big\}.
\end{eqnarray}
Note that the deepest bracket 
$\displaystyle\Big\{\big|(\partial_{\th}^{n-p}\Psi)(2^j t-k,\th)\big|+ \big|(\partial_{\th}^{n-p}\Psi)(-k,\th)\big|\Big\}$  contains two  terms: the first
$\displaystyle \big|(\partial_{\th}^{n-p}\Psi)(2^j t-k,\th)\big|$
depends on $t\in{\cal K}$ whilst the second $\displaystyle  \big|(\partial_{\th}^{n-p}\Psi)(-k,\th)\big|$ no longer depends on $t$. Therefore, it suffices to obtain a bound of the supremum for $t\in{\cal K}$ of the sum corresponding to the first term,  then to   use it in the special case ${\cal K}=\{ 0\}$ to bound the sum corresponding to the second term.
Let us  remark that there exists a real $K>0$ such that ${\cal K}\subset
[-K,K]$. Thus,  without any restriction, we can suppose that ${\cal
  K}=[-K,K]$. Next, using (\ref{eq:uniflocpsi}), the convention that $0^0=1$, the change of variable $k=k'+[2^jt]$, the fact that $|t|\le K$, (\ref{bound:log3xy}) and the fact that
$\displaystyle z=2^jt-[2^jt]\in[0,1]$ , one has the following estimates for each $p\in\{0,\ldots,n\}$ and $(t,\th)\in [-K,K]\times [a,b]$:
\begin{eqnarray}
\label{eq2:boundg}
\nonumber
&& \sum_{j=0}^{+\infty}\sum_{k=-\infty}^{+\infty} j^p 2^{-j\th}\,\sqrt{\log (3+j+|k|)}\,\big|(\partial_{\th}^{n-p}\Psi)(2^j t-k,\th)\big|\\
\nonumber
&& \le c_2\sum_{j=0}^{+\infty}\sum_{k=-\infty}^{+\infty} j^p 2^{-ja}\sqrt{\log(3+j+|k|)}\cdot(2+|2^jt-k|)^{-\ell}
\nonumber
\\
&& \le c_2\sum_{j=0}^{+\infty}\sum_{k'=-\infty}^{+\infty} j^p 2^{-ja}\sqrt{\log(3+j+|k'|+2^j K)}\cdot(2+|2^jt-[2^jt]-k'|)^{-\ell}
\nonumber
\\
&& \le c_2 c_3\sum_{j=0}^{+\infty}j^p 2^{-ja}\sqrt{\log(3+j+2^j K)}<\infty,
\end{eqnarray}
where
\begin{equation}
\label{eq3:boundg}
c_3=\sup\Big\{\sum_{k=-\infty}^{+\infty}(2+|z-k|)^{-l}\sqrt{\log(3+|k|)}:\,\,z\in [0,1]\Big\}<\infty.
\end{equation}
Clearly, (\ref{eq2:boundg}) combined with (\ref{maj:An}) implies that (\ref{eq1:boundg}) holds.
\Box\par\bigskip\noindent

{\it Proof of Lemma \ref{lem:d1boundg}.} The proof is very technical so let us
first explain the main ideas behind it. For the sake of simplicity, we make
the change of notation  $t_1=t_0+h$. Then we split the set of indices
$\{(j,k)\in \nit\times\zit\}$ into three disjoint subsets: ${\cal V}$
a neighborhood of radius $r$ about $t_0$, a subset ${\cal W}$ corresponding to the the low frequency ($j\le j_1$) outside the neighborhood ${\cal V}$ and a subset
${\cal W}^c$ corresponding to the the high frequency ($j> j_1$) outside the
neighborhood $ {\cal V}$ (the ``good'' choices of the radius $r$ and of the
cutting frequency $j_1$ will be clarified soon). Thus the sum through which
$G_n(t_0,t_1,\th)$ is defined (see the statement of Lemma \ref{lem:d1boundg})
can be decomposed into three parts: a sum over ${\cal V}$, a sum over ${\cal
  W}$ and a sum over ${\cal W}^c$; they respectively be denoted
$B_{1,n}(t_0,h,\th)$, $B_{2,n}(t_0,h,\th)$ and $B_{3,n}(t_0,h,\th)$. In order
to be able to show that, to within a constant, each of these three quantities
is upper bounded by $|h|^{d_n}$ for some exponent $d_n>b$, we need to
conveniently choose the radius $r$ of the neighborhood ${\cal V}$ as well as
the cutting frequency $j_1$. The most natural choice is to take $r=|h|$ and
$2^{-j_1}\simeq |h|$. However a careful inspection of the proof of Lemma
\ref{lem:d4boundg} shows this does not work basically because $2^{j_1}|h|$
does not go to infinity when $|h|$ tends $0$. Roughly speaking, to overcome
this difficulty we have taken $r=|h|^{\eta}$ and  $2^{-j_1}\simeq
|h|^{\gamma}$ where $0<\eta<\gamma<1$ are two parameters (the ``good'' choices
of these parameters will be clarified soon)
as shown by  the following Figure
\par
\vspace{-0.2cm}
\begin{center}
\includegraphics[width=10cm,height=6.5cm]{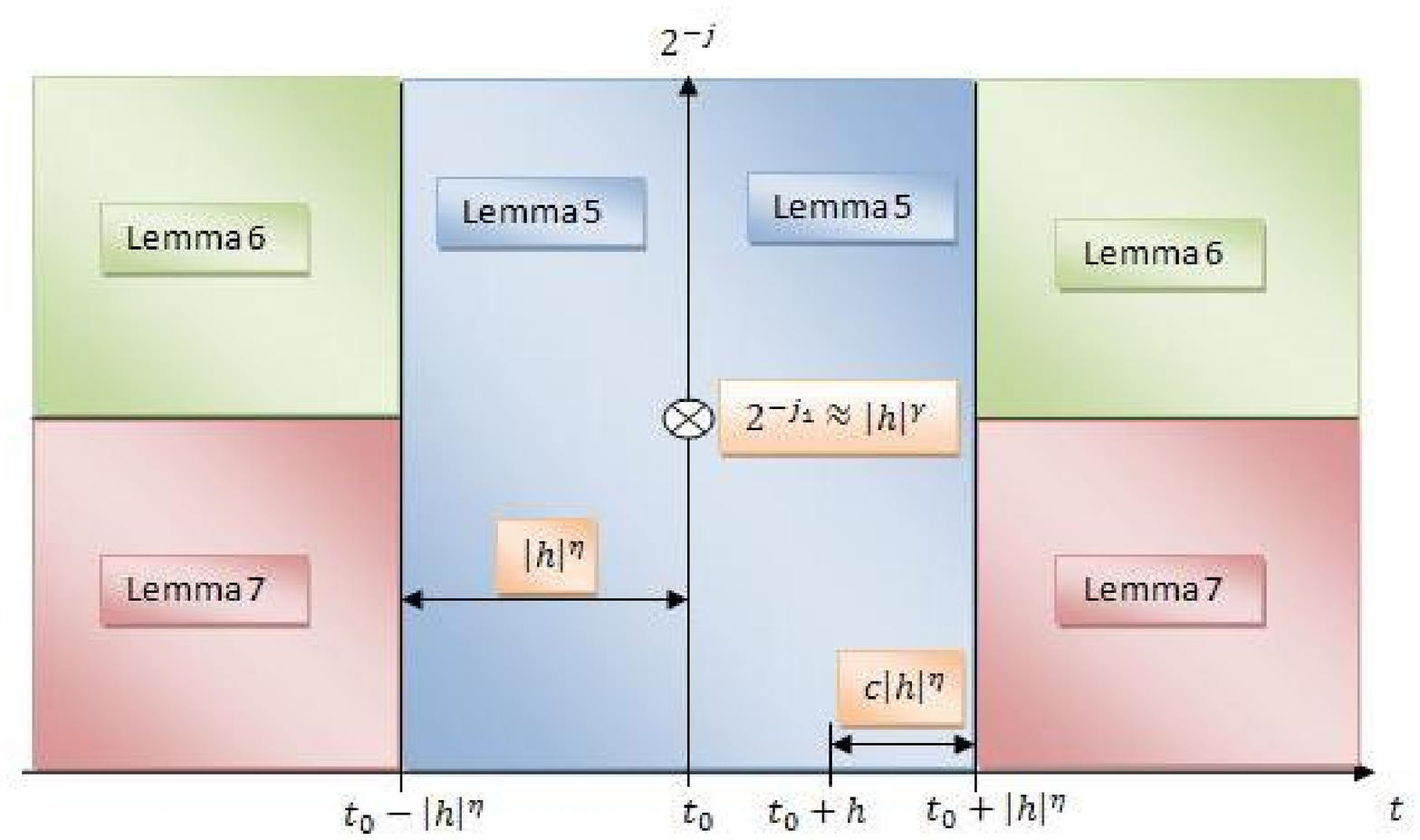}
\\
{\bf Fig.2,} {\em  the three Lemmas and the corresponding subset of indices}
\end{center}
More precisely, $j_1$ is the unique nonnegative integer satisfying
\begin{equation}
\label{eq:defj1}
2^{-j_1-1}<|h|^\gamma \le 2^{-j_1}
\end{equation}
and the sets ${\cal V}$, ${\cal W}$ and ${\cal W}^c$ are defined by:
\begin{equation}
\label{eq:defVt0}
 {\cal V}(t_0,h,\eta)=\{(j,k)\in\nit\times\zit :\,\,|k/2^j-t_0|\le |h|^\eta\},
\end{equation}
\begin{equation}
\label{eq:defVt0c}
 {\cal V}^c(t_0,h,\eta)=\{(j,k)\in\nit\times\zit :\,\,|k/2^j-t_0|> |h|^\eta\},
\end{equation}
\begin{equation}
\label{eq:defWt0}
{\cal W}(t_0,h,\eta,\gamma)=\{(j,k)\in {\cal V}^c(t_0,h,\eta):\,\,0\le j\le j_1\}
\end{equation}
and
\begin{equation}
\label{eq:defWt0c}
{\cal W}^c(t_0,h,\eta,\gamma)=\{(j,k)\in {\cal V}^c(t_0,h,\eta):\,\, j\ge j_1+1\}.
\end{equation}
It follows from Lemmas \ref{lem:leibniz} to \ref{lem:d4boundg} that
\begin{eqnarray}
\label{eq2:d1boundg}
\nonumber
&& G_n(t_0,t_0+h,\th)=\sum_{m=1}^3 B_{m,n}(t_0,h,\th)\\
&&\le \sum_{p=0}^n \sum_{m=1}^3 C_n^p (\log 2)^p B_{m,n,p}(t_0,h,\th)\\
\nonumber
&& \le c_{4}\Big (|h|^{a+\eta\beta}+|h|^{(1-\gamma)+\gamma a}+|h|^{(\gamma-\eta)(\ell-1-\eps)+\gamma a}\Big )\log^{n+1/2}(1/|h|),
\end{eqnarray}
where the constant $c_{4}=\max\{c_{5}, c_{8},c_{10}\} \sum_{p=0}^n C_n^p (\log 2)^p$ does not depend on $(t_0,h,\th)$.
In view of (\ref{eq2:d1boundg}) and the inequality $\beta>b$ as well as the fact that $\eps$ is arbitrarily small, for proving that (\ref{eq1:d1boundg}) holds its sufficient to show that there exist two reals $0<\eta<\gamma<1$ and an integer $l\ge 2$ satisfying the following inequalities
$$
\left\{
\begin{array}{l}
 a+\eta b\ge b\\
(1-\gamma)+\gamma a\,>b\\
(\gamma-\eta)(\ell-1)+\gamma a\,>\,b.
\end{array}
\right.
$$
This is clearly the case. In fact, thanks to (\ref{eq:condab}) we can easily
show that the first two inequalities have common solutions; moreover each of
their common solutions is also a solution of the third inequality provided that $\ell$ is big enough.
\Box\par\medskip\noindent

Before ending this section let us prove that Corollary~\ref{cor:main2} holds.
\par
\medskip
\noindent
{\it Proof of Corollary \ref{cor:main2}.} Let us first show that $Z$ has the same feature $(b)$ as mBm. In view of Theorem \ref{thm:main1} and Remark \ref{rem:cond19} it is clear that $\a_R$, the pointwise H\"older exponent of $R$, satisfies a.s. for all $t\in\rit$,
\begin{equation}
\label{eq:lbpheR}
\a_R (t)\ge d.
\end{equation}
Next putting together (\ref{eq:lbpheR}), the fact that $d>b$, (\ref{eq:defR}) and (\ref{eq:phembm}) it follows that a.s. for all $t\in\rit$,
$$
\a_Z (t)=H(t).
$$
Let us now show that $Z$ has the same feature $(c)$ as mBm. Let $(\rho_n)$ be an arbitrary sequence of positive real numbers converging to $0$. In view of (\ref{eq:defR}) and (\ref{eq:tangentX}), to prove that for each $t\in\rit$ one has
\begin{equation}
\label{eq:tangentZ}
\lim_{n\rightarrow\infty}\mbox{law}\left\{\frac{Z(t+\rho_n u)-Z(t)}{\rho_n^{H(t)}}:\,\,u\in\rit\right\}=\mbox{law}\{B_{H(t)}(u):\,\,u\in\rit\},
\end{equation}
in the sense of finite dimensional distribution, it is sufficient to prove that for any $u\in\rit$ one has
\begin{equation}
\label{eq:zerotangentR}
\lim_{n\rightarrow +\infty} E\left\{\Big (\frac{R(t+\rho_n u)-R(t)}{\rho_n^{H(t)}}\Big)^2\right\}=0.
\end{equation}
Observe that for all $n$ big enough one has $\rho_n|u|\le 1$. Therefore, taking  ${\cal K}=[t-1,\, t+1]$ in Theorem \ref{thm:main1}, it follows that for $n$ big enough,
\begin{equation}
\label{eq:inegtangR}
 E\left\{\Big (\frac{R(t+\rho_n u)-R(t)}{\rho_n^{H(t)}}\Big)^2\right\}\le \rho_{n}^{2(d-H(t))} E(C_{1}^{2})
\end{equation}
and the latter inequality clearly implies that (\ref{eq:zerotangentR}) holds. To have in (\ref{eq:tangentZ}) the convergence in distribution for the topology of the uniform convergence on compact sets it is sufficient to show that for any positive real $L$, the sequence of continuous Gaussian processes,
$$
\left\{\frac{Z(t+\rho_n u)-Z(t)}{\rho_n^{H(t)}}:\,\,u\in [-L,L]\right\},\, n\in\nit,
$$
is tight. This tightness result can be obtained (see \cite{b68}) by proving that there exists a constant $c_{17}>0$ only depending on $L$ and $t$ such that for all $n\in\nit$ and each $u_1,u_2\in [-L,L]$ one has
\begin{equation}
\label{eq:tight}
E\left\{\Big (\frac{Z(t+\rho_n u_1)-Z(t)}{\rho_n^{H(t)}}-\frac{Z(t+\rho_n u_2)-Z(t)}{\rho_n^{H(t)}}\Big )^2\right\}\le c_{17}|u_1-u_2|^{2H(t)}.
\end{equation}
Without loss of generality we may assume that for every $n\in\nit$, $\rho_n\in
(0,1]$. Then by using the fact that (\ref{eq:tight}) is satisfied when $Z$ is
replaced by $X$ (see \cite{bci98} Proposition 2) as well as the fact that it is also satisfied when $Z$ is replaced by $R$ (this can be done similarly to (\ref{eq:inegtangR})), one can establish that this inequality holds.
\Box\par\medskip\noindent

\section{Some technical Lemmas}

\begin{lemma}
\label{lem:leibniz}
 For every integer $n\ge 0$ and any $(t,\th)\in\rit\times\rit$ one has
\begin{eqnarray}
\label{eq1:leibniz}
&&(\partial_{\th}^n g_{j,k})(t,\th)\\
\nonumber
&& =\sum_{p=0}^n C_{n}^p (-j\log 2)^p 2^{-j\th}\Big\{(\partial_{\th}^{n-p}) \Psi (2^jt-k,\th)-(\partial_{\th}^{n-p} \Psi)(-k,\th)\Big\}.
\end{eqnarray}
\end{lemma}
{\it Proof of Lemma \ref{lem:leibniz}.}  The lemma can easily be obtained by applying
the Leibniz formula for the $n$th derivative of a product of two functions.
\Box\par\medskip\noindent

\begin{lemma}
\label{lem:d2boundg}
For each integer $n\ge 1$ and $(t_0,h,\th)\in\rit\times \rit \times (0,+\infty)$ set
\begin{eqnarray*}
B_{1,n,p}(t_0,h,\th)& :=&\hspace{-0.3cm}\sum_{(j,k)\in {\cal V}(t_0,h,\eta)} j^p 2^{-j\th} |H(t_0)-H(k/2^j)|^n\times\sqrt{\log(3+j+|k|)}
\\
&&\hspace{0.5cm}\times
\Big|(\partial_\th^{n-p}\Psi)(2^j(t_0+h)-k,\th)-(\partial_\th^{n-p}\Psi) (2^j t_0-k,\th)\Big|,
\end{eqnarray*}
where ${\cal V}(t_0,h,\eta)$ is the set defined by (\ref{eq:defVt0}).
Then, for all real $K>0$ and every integers $n\ge 1$ and $0\le p\le n$, one has
\begin{eqnarray}
\label{eq1:d2boundg}
\nonumber
c_5:=\sup_{(t_0,\th)\in [-K,K]\times [a,b], |h|<1/4} |h|^{-a-n\eta\beta}\log^{-p-1/2}(1/|h|)B_{1,n,p}(t_0,h,\th)<\infty.\\
\end{eqnarray}
\end{lemma}
{\it Proof of Lemma \ref{lem:d2boundg}.} It follows from (\ref{eq:HoldH}) and (\ref{eq:defVt0}) that
\begin{eqnarray}
\label{eq2:d2boundg}
&& \sum_{(j,k)\in {\cal V}(t_0,h,\eta)} j^p 2^{-j\th} |H(t_0)-H(k/2^j)|^n\times\sqrt{\log(3+j+|k|)}
\\
\nonumber&&
\hspace{3cm}\times\Big|(\partial_\th^{n-p}\Psi)(2^j(t_0+h)-k,\th)-(\partial_\th^{n-p}\Psi)(2^j t_0-k,\th)\Big|
\\
\nonumber
&& \le c_1|h|^{n\beta\eta}\sum_{(j,k)\in {\cal V}(t_0,h,\eta)} j^p 2^{-ja} \times\sqrt{\log(3+j+|k|)}
\\
\nonumber&&
\hspace{3cm}\times\Big|(\partial_\th^{n-p}\Psi)(2^j(t_0+h)-k,\th)-(\partial_\th^{n-p}\Psi)(2^j t_0-k,\th)\Big|.
\end{eqnarray}
Now let $j_0\ge 2$ be the unique integer such that
\begin{equation}
\label{eq:defj0}
2^{-j_0-1}<|h|\le 2^{-j_0}.
\end{equation}
By using (\ref{eq:uniflocpsi}), (\ref{eq:defj0}), the change of variable
$k=k'+[2^j y]$,   (\ref{bound:log3xy}) and the fact that $|t_0|\le K$, we can deduce that for any $y\in [t_0-1,t_0+1]$,
\begin{eqnarray}
\label{eq3:d2boundg}
\nonumber
&& \sum_{j=j_0+1}^{+\infty}\sum_{k=-\infty}^{+\infty}j^p 2^{-ja}|(\partial_\th^{n-p}\Psi)(2^j y-k,\th)|\sqrt{\log(3+j+|k|)}\\
\nonumber
&& \le c_2\hspace{-0.1cm}\sum_{j=j_0+1}^{+\infty}\sum_{k'=-\infty}^{+\infty}j^p 2^{-ja}(2+|2^j y-[2^j y]-\hspace{-0.1cm}k'|)^{-\ell}\sqrt{\log(3\hspace{-0.1cm}+j\hspace{-0.1cm}+2^j(|t_0|\hspace{-0.1cm}+\hspace{-0.1cm}1)\hspace{-0.1cm}+\hspace{-0.1cm}|k'|)}\\
\nonumber
&& \le c_2\cdot c_3\sum_{j=j_0+1}^{+\infty}j^p 2^{-ja}\sqrt{\log(3+j+2^j(K+1))}\\
&& \le c_6 |h|^{a} \log^{p+1/2}(1/|h|),
\end{eqnarray}
where $c_3$ is the  constant defined by (\ref{eq3:boundg}) and the last
inequality (in which $c_6$ is a constant non depending on $(t_0,h,\eta)$) follows from (\ref{eq:defj0}) and some classical and easy calculations.

On the other hand, by using the Mean-value Theorem applied to the function
$\partial_\th^{n-p}g_{j,k}$ with respect to the first variable,
(\ref{eq:uniflocpsi}), the fact that for all $2^j|h|\le 1$ for all
$j\in\{0,\ldots, j_0\}$,  (\ref{eq:defj0}), (\ref{bound:log3xy}),
(\ref{eq3:boundg}) and the inequality $|t_0|\le K$, we get that
\begin{eqnarray}
\label{eq4:d2boundg}
\nonumber
&& \sum_{j=0}^{j_0}\sum_{k=-\infty}^{+\infty} j^p 2^{-ja}\times \sqrt{\log(3+j+|k|)}\\ &&\hspace{3cm}\times\Big|(\partial_\th^{n-p}\Psi)(2^j(t_0+h)-k,\th)-(\partial_\th^{n-p}\Psi) (2^j t_0-k,\th)\Big|
\nonumber
\\
\nonumber
&& \le c_2 |h|\sum_{j=0}^{j_0}\sum_{k=-\infty}^{+\infty} j^p 2^{j(1-a)}(1+|2^j t_0-[2^j t_0]-k|)^{-\ell}\sqrt{\log(4+j+2^j|t_0|+|k|)}\\
\nonumber
&& \le c_2 c_3 |h|\sum_{j=0}^{j_0} j^p 2^{j(1-a)}\sqrt{\log(4+j+2^j K)}\\
&& \le c_7 |h|^{a} \log^{p+1/2}(1/|h|),
\end{eqnarray}
where the constant $c_7$ does not depend on $(t_0,h,\eta)$.
Finally, by combining  (\ref{eq2:d2boundg}) with (\ref{eq3:d2boundg}) and (\ref{eq4:d2boundg}), one can deduce
(\ref{eq1:d2boundg}).
\Box

\begin{lemma}
\label{lem:d3boundg}
For any $(t,h,\th)\in\rit\times \rit \times (0,+\infty)$ and for any integers $n\ge 1$ and $0\le p\le n$ set
\begin{eqnarray*}
B_{2,n,p}(t_0,h,\th)& :=&\hspace{-0.2cm}\sum_{(j,k)\in {\cal W}(t_0,h,\eta,\gamma)}\hspace{-0.2cm} j^p 2^{-j\th}   |H(t_0)-H(k/2^j)|^n\times\sqrt{\log(3+j+|k|)}
\\
&&
\quad\times
\Big|(\partial_\th^{n-p}\Psi)(2^j(t_0+h)-k,\th)-(\partial_\th^{n-p}\Psi) (2^j t_0-k,\th)\Big|,
\end{eqnarray*}
where ${\cal W}(t_0,h,\eta,\gamma)$ is the set defined by (\ref{eq:defWt0}). Then, for any real $K>0$, one has that
\begin{eqnarray}
\nonumber
c_8&=&\hspace{-0.3cm}\sup_{(t_0,\th)\in [-K,K]\times [a,b], |h|<1/4}|h|^{-(1-\gamma)-\gamma a}\log^{-p-1/2}(1/|h|)B_{2,n,p}(t_0,h,\th)
\\
\label{eq1:d3boundg}
&&\hspace{9cm}<\infty.\hspace{1cm}
\end{eqnarray}
\end{lemma}
{\it Proof of Lemma \ref{lem:d3boundg}}. To begin with, note that for any pair of  real numbers $(\th_0,\th_1)\in (0,1)^2$, one has $|\th_1-\th_0| <1$. Therefore,
\begin{equation}\label{bound:H}
\mathrm{for\; all}\;(j,k) \in \nit\times \zit,\qquad |H(t_0)-H(k/2^j)|^n<1.
\end{equation}
By using the Mean-value Theorem applied to the function $\partial_\th^{n-p}\Psi$ with respect to the first variable combined with (\ref{eq:uniflocpsi}), we get  for all $t_0\in \cal K$ and $h\in \rit$
\begin{eqnarray*}
\Big|(\partial_\th^{n-p}\Psi)(2^j(t_0+h)-k,\th)-(\partial_\th^{n-p}\Psi) (2^j t_0-k,\th)\Big|\hspace{3cm}&&\\
\le
c_2\, 2^j |h|\,\big(2+|2^j t_0-k+2^j uh |\big)^{-\ell} &&
\end{eqnarray*}
for a real number $u\in (0,1)$. On the other hand it follows the inequality $2^j |h| \le 1$ for all $j\in\{0,\ldots, j_1\}$ (which is a consequence of (\ref{eq:defj1})) and from triangle inequality that $|2^j t_0-k+2^j uh |\ge |2^j t_0-k|-1$. Therefore
\begin{eqnarray*}
\Big|(\partial_\th^{n-p}\Psi)(2^j(t_0+h)-k,\th)-(\partial_\th^{n-p}\Psi) (2^j t_0-k,\th)\Big|\hspace{3cm}&&\\
\le
c_2\, 2^j |h|\,\big(1+|2^j t_0-k|\big)^{-\ell} &&
\end{eqnarray*}
and as a consequence, we obtain for all  $(t,h,\th)\in\rit\times \rit \times (0,1)$
\begin{eqnarray*}
\label{eq2:d3boundg}
B_{2,n,p}(t_0,h,\th)
&& \le c_2 |h|\sum_{j=0}^{j_1}\sum_{k=-\infty}^{+\infty} j^p\, 2^{j(1-\th)}(1+|2^j t_0-k|)^{-\ell}\sqrt{\log(3+j+|k|)}.
\end{eqnarray*}
Next, making the change of variable $k=k'+ [2^j t_0]$ and using triangle inequality as well as the inequality $\th\ge a$,
we deduce that for all $(t,h,\th)\in[-K,K]\times[-1/4,1/4]\times [a,b]$
\begin{eqnarray*}
\label{eq2:d3boundg}
B_{2,n,p}(t_0,h,\th)
&\le& c_2 |h|\sum_{j=0}^{j_1}\sum_{k'=-\infty}^{+\infty} j^p 2^{j(1-a)}(1+|2^j t_0-[2^j t_0]-k'|)^{-\ell}\\
\nonumber&&
\hspace{5cm}\times\sqrt{\log(3+j+2^{j}|t_0|+|k'|)}
\\
&\le&
 c_2 |h|\left\{\sum_{k'=-\infty}^{+\infty} \sqrt{\log(3+|k'|)}(1+|2^j t_0-[2^j t_0]-k'|)^{-\ell}\right\}
\\
&&
\hspace{0.5cm}
\times
 \left\{\sum_{j=0}^{j_1}j^p 2^{j(1-a)}\sqrt{\log(4+j+2^{j}K|)}\right\},
\end{eqnarray*}
where the last inequality follows from $|t_0|\le K$ and the inequality (\ref{bound:log3xy}).
Set $z=2^j t_0-[2^j t_0]$, obviously $z\in [0,1)$, thus (\ref{eq3:boundg}) and the latter inequality imply that
\begin{eqnarray*}
\label{eq2:d3boundg}
B_{2,n,p}(t_0,h,\th)
&\le&
 c_2\cdot c_3\, |h|\,
 \left\{\sum_{j=0}^{j_1}j^p 2^{j(1-a)}\sqrt{\log(4+j+2^{j}K|)}\right\},
\end{eqnarray*}
for all $(t,h,\th)\in[-K,K]\times [-1/4,1/4]\times [a,b]$. Finally, in view of the inequalities $2^{j_1}\le |h|^{-\gamma}$ and $j_1\le \log(1/|h|)$ (these inequalities are a consequences of (\ref{eq:defj1})), we get
\begin{eqnarray}
\label{eq2:d3boundg}
B_{2,n,p}(t_0,h,\th)\le c_9 |h|^{(1-\gamma)+\gamma a}\log^{p+1/2}(1/|h|),
\end{eqnarray}
where the constant $c_9$ does not depend on $(t_0,h,\th)$.
This finishes the proof of Lemma \ref{lem:d3boundg}.
\Box\par\medskip\noindent

\begin{lemma}
\label{lem:d4boundg}
For any $(t,h,\th)\in\rit\times \rit \times [a,b]$ and any integers $n\ge 1$ and $0\le p\le n$ set
\begin{eqnarray*}
B_{3,n,p}(t_0,h,\th)& :=&\hspace{-0.3cm}\sum_{(j,k)\in {\cal W}^c(t_0,h,\eta,\gamma)} \hspace{-0.2cm} j^p 2^{-j\th}   |H(t_0)-H(k/2^j)|^n\times\sqrt{\log(3+j+|k|)}
\\
&&
\quad\times
\Big|(\partial_\th^{n-p}\Psi)(2^j(t_0+h)-k,\th)-(\partial_\th^{n-p}\Psi) (2^j t_0-k,\th)\Big|
\end{eqnarray*}
where ${\cal W}^c(t_0,h,\eta,\gamma)$ is the set defined by (\ref{eq:defWt0c}). Then, for every real $K>0$, for each arbitrarily small real $\eps >0$ and all integer $l\ge 2$, one has $c_{10}<\infty$ where
\begin{eqnarray*}
\label{eq1:d4boundg}
c_{10}&:=&\hspace{-0.2cm}\sup_{(t_0,\th)\in [-K,K]\times [a,b], |h|<1/4}\hspace{-0.2cm}|h|^{-(\gamma-\eta)(l-1-\eps)-\gamma a}\log^{-p-1/2}(|h|^{-1})B_{2,n,p}(t_0,h,\th)
\end{eqnarray*}
\end{lemma}
\par
\medskip
\noindent
{\it Proof of Lemma \ref{lem:d4boundg}.}
By using the triangle inequality combined with (\ref{eq:defWt0c}) and (\ref{eq:defVt0c}), one gets,
for all $(j,k)\in {\cal W}^c(t_0,h,\eta,\gamma)$,
\begin{eqnarray}
\label{eq2:d4boundg}
|(t_0+h)-k 2^{-j}|&\ge& |t_0-k 2^{-j}+h|\ge  |h|^{\eta}-|h|\ge c_{11}|h|^{\eta},
\end{eqnarray}
where the constant $c_{11}=1-4^{\eta-1}$. This means that the integer $k$ necessarily satisfies
\begin{eqnarray}
\label{eq5:d4boundg}
|2^j(t_0+h)-k |\ge c_{11}2^j|h|^{\eta}.
\end{eqnarray}
In view of (\ref{eq5:d4boundg}), let us consider ${\cal T}_{j}^{+}$ and ${\cal T}_{j}^{-}$ the sets of positive real numbers defined by
$$
{\cal T}_{j}^{+}=\{|2^j(t_0+h)-k |:\,\,\mbox{ $k\in\zit$ and $2^j(t_0+h)- k\ge c_{11}2^j|h|^{\eta}$}\}
$$
and
$$
{\cal T}_{j}^{-}=\{|2^j(t_0+h)-k |:\,\,\mbox{ $k\in\zit$ and $k-2^j(t_0+h)\ge c_{11}2^j|h|^{\eta}$)}\}.
$$
For every fixed $j$, the set ${\cal T}_{j}^{+}$ can be viewed as a strictly increasing sequence $(\tau_{j,q}^{+})_{q\in\nit}$ satisfying for all $q\in\nit$,
\begin{equation}
\label{eq6:d4boundg}
q+c_{11}2^j|h|^{\eta}\le \tau_{j,q}^{+}<  q+1+c_{11}2^j|h|^{\eta}.
\end{equation}
Similarly, for every fixed $j$, the set ${\cal T}_{j}^{-}$ can be viewed as a strictly increasing sequence $(\tau_{j,q}^{-})_{q\in\nit}$ satisfying for all $q\in\nit$,
\begin{equation}
\label{eq7:d4boundg}
q+c_{11}2^j|h|^{\eta}\le\tau_{j,q}^{-}<q+1+c_{11}2^j|h|^{\eta}.
\end{equation}
Next, setting ${\cal T}_j={\cal T}_{j}^{+}\cup{\cal T}_{j}^{-}$, it follows that from (\ref{eq:uniflocpsi}), the triangle inequality, the inequality $|t_0+h|\le K+1$, (\ref{eq:defWt0c}), (\ref{eq5:d4boundg}), (\ref{eq6:d4boundg}) and (\ref{eq7:d4boundg}) that
\begin{eqnarray}
\label{eq3:d4boundg}
\nonumber
&& \sum_{(j,k)\in {\cal W}^c(t_0,h,\eta,\gamma)}j^p 2^{-j\th}\Big|(\partial_\th^{n-p}\Psi)(2^j(t_0+h)-k,\th)\Big|\sqrt{\log(3+j+|k|)}\\
\nonumber
&& \le c_2 \hspace{-0.7cm}\sum_{(j,k)\in {\cal W}^c(t_0,h,\eta,\gamma)}j^p 2^{-j\th}\Big(2+|2^j(t_0+h)-k |\Big )^{-\ell}\times\dots\\
\nonumber
&& \hspace{4cm}\times\sqrt{\log(3+j+|2^j(t_0+h)-k |+2^j(K+1))}\\
\nonumber
&& \le c_2 \hspace{-0.2cm}\sum_{j=j_1+1}^{+\infty}\sum_{\tau\in{\cal T}_j} j^p 2^{-j\th}\Big(2+|\tau|\Big )^{-\ell}
\sqrt{\log(3+j+|\tau|+2^j(K+1))}\\
\nonumber
&& \le 2 c_2 \hspace{-0.2cm}\sum_{j=j_1+1}^{+\infty}\sum_{q=0}^{+\infty}j^p 2^{-ja}
\Big(2+q\hspace{-0.1cm}+c_{11}2^j|h|^{\eta} \Big )^{-\ell}\times\dots\\
\nonumber
&& \hspace{4cm}\times\sqrt{\log(4+j+q+2^j|h|^{\eta}+2^j(K+1))}
\end{eqnarray}
Then, one can use the inequality $\displaystyle (2+x)^{-\ell}\sqrt{\log(4+x)}\le c_{12}(2+x)^{-\ell+\eps}$ which is  valid for all nonnegative real number $x$ where $\eps$ is a fixed arbitrarily small positive real number and $c_{12}$ is a constant only depending on $\eps$. By combining this inequality with (\ref{bound:log3xy}), (\ref{eq:defj1}) and  $|h|\le 1/4$, we get
\begin{eqnarray}
\label{eq3:d4boundg}
\nonumber
&& \sum_{(j,k)\in {\cal W}^c(t_0,h,\eta,\gamma)}j^p 2^{-j\th}\Big|(\partial_\th^{n-p}\Psi)(2^j(t_0+h)-k,\th)\Big|\sqrt{\log(3+j+|k|)}\\
\nonumber
&& \le 2 c_2 \sum_{j=j_1+1}^{+\infty}\sum_{q=0}^{+\infty}j^p 2^{-ja}
\sqrt{\log(3+j+2^j(K+1))}\times\dots\\
\nonumber
&&
\hspace{2cm}\times\Big(2\hspace{-0.1cm}+q\hspace{-0.1cm}+ c_{11}2^j|h|^{\eta}\Big )^{-\ell}\sqrt{\log(4+q+c_{11}2^j|h|^{\eta})}\\
\nonumber
&&  \le 2 c_2 c_{12}\sum_{j=j_1+1}^{+\infty}j^p 2^{-ja} \sqrt{\log(3+j+2^j(K+1))}
\times\dots\\
\nonumber
&& \hspace{5cm}
\times\left(\int_{0}^{+\infty}\hspace{-0.1cm}\Big(1\hspace{-0.1cm}+y\hspace{-0.1cm}+c_{11}2^j|h|^{\eta}\Big )^{-\ell+\eps}dy\right)\\
\nonumber
&& \le c_{13}\sum_{j=j_1+1}^{+\infty}j^p 2^{-ja} \sqrt{\log(3+j+2^j(K+1)))}
\Big(1+ c_{11}2^j|h|^{\eta}\Big )^{-(\ell-1-\eps)}\\
\nonumber
&& \le c_{14}|h|^{-\eta(\ell-1-\eps)}\sum_{j=j_1+1}^{+\infty}j^p
2^{-j(a+\ell-1-\eps)} \sqrt{\log(3+j+2^j(K+1))}\\
\nonumber
&& \le c_{15}\, |h|^{(\gamma-\eta)(\ell-1-\eps)+\gamma a}\log^{p+1/2}(1/|h|),
\end{eqnarray}
where $c_{12}$, $c_{13}$, $c_{14}$ and $c_{15}$ are constants which do not
depend on $(t_0,h,\th)$. Finally, using the latter inequality as well as the
triangle inequality and the
fact that $H(\cdot)$ is with values in $[a,b]$ we get the lemma.
\Box
\\
\\
{\bf Acknowledgments.}  The authors thank the anonymous referee for many useful remarks
which greatly helped them to improve the earlier version of this article.

\input{referenc}



\end{document}

%% file: referenc.tex
%
%
%

%
%